\pdfoutput=1
\documentclass[aps,prl,twocolumn,superscriptaddress,longbibliography,floatfix]{revtex4-2}

\usepackage{amsmath,amssymb}
\usepackage{graphicx}
\usepackage{bm}
\usepackage[colorlinks=true,linkcolor=blue,citecolor=blue,urlcolor=blue]{hyperref}
\usepackage{xcolor}
\newcommand{\cptitlerevision}[1]{#1}
\graphicspath{{figs/}}

\makeatletter
\let\frontmatter@footnote@produce\frontmatter@footnote@produce@footnote
\makeatother

\newcommand{\Om}{\Omega}

\begin{document}

\title{Magnetizing nonlinear plasma wakefields for positron acceleration}

\author{Yung-Kun Liu}
\email[Corresponding author: ]{r06222017@ntu.edu.tw}
\affiliation{Leung Center for Cosmology and Particle Astrophysics (LeCosPA),
National Taiwan University, Taipei 10617, Taiwan}
\affiliation{Department of Physics, National Taiwan University, Taipei 10617, Taiwan}
\author{Pisin Chen}
\affiliation{Leung Center for Cosmology and Particle Astrophysics (LeCosPA),
National Taiwan University, Taipei 10617, Taiwan}
\affiliation{Department of Physics, National Taiwan University, Taipei 10617, Taiwan}
\author{Ching-En Lin}
\affiliation{Department of Applied Physics, Stanford University, Stanford,
California 94305, USA}
\affiliation{SLAC National Accelerator Laboratory, Menlo Park, California 94025, USA}
\author{Spencer Gessner}
\affiliation{SLAC National Accelerator Laboratory, Menlo Park, California 94025, USA}
\author{Bernhard Hidding}
\affiliation{Department of Physics, Heinrich Heine University, D\"{u}sseldorf, Germany}

\begin{abstract}
It is known that only a narrow plasma wakefield sliver in the electron-beam-driven blowout regime suits positron acceleration. Using 3D simulations, we show that matching the cyclotron frequency $\omega_{c}$ with the plasma frequency $\omega_{p}$ forms a stable electron column on axis, expanding the suitable phase space for positron acceleration sizably. For a plasma density $n_p = 10^{16}\text{ cm}^{-3}$ in a 35 T field, the interval expands 4.3 times, and a witness positron beam gains 100--150 MeV over 6 cm ($1.6$--$2.5$~GeV/m) with a 92\% capture rate.
\end{abstract}

\maketitle

Plasma wakefield accelerators are able to sustain accelerating gradients orders of magnitude beyond radio-frequency technology~\cite{Tajima1979,Chen1985}. However, the nonlinear blowout regime~\cite{Rosenzweig1991,Lu2006} that serves as the workhorse for electron acceleration fails for positrons because the ion cavity is entirely defocusing. Positrons can be accelerated and focused together only in a narrow, highly nonlinear region where the sheath crosses the axis. This ``positron problem" remains a major hurdle for a plasma-based $e^+e^-$ collider~\cite{Cao2024,Joshi2025,ChenLiu2026}. Representative remedies include re-engineering the wakefield via hollow channels~\cite{Gessner2016,Lindstrom2018,Silva2021}, plasma columns~\cite{Diederichs2019,Diederichs2020,Diederichs2022}, or shaped drivers~\cite{Wang2021,Jain2015,Vieira2014} to provide transverse forces positrons can survive. Yet, each requires tailored plasmas, auxiliary beams, or complex drivers. Positron-driven wakefields have also been used to accelerate positrons experimentally~\cite{Corde2015,Doche2017}, where the trailing bunch loads and reshapes the wake it rides on~\cite{Zhou2025}.

\begin{figure*}[t]
\includegraphics[width=0.8\textwidth]{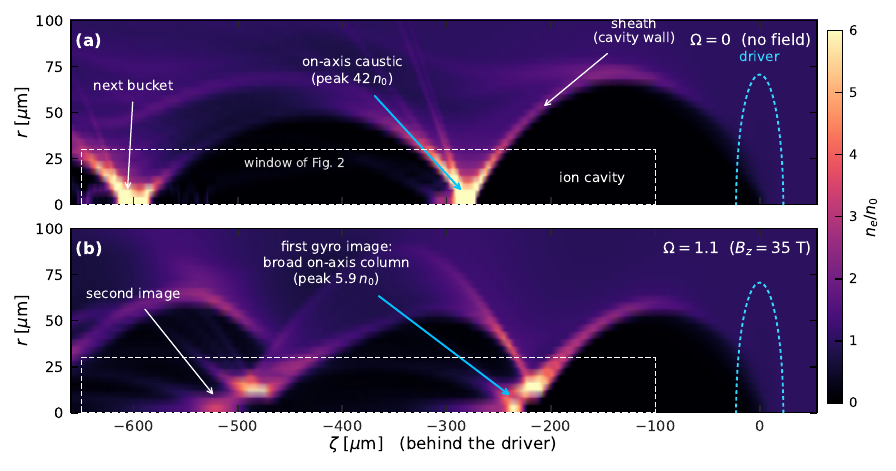}
\caption{\label{fig:bubble}
Impact of the axial magnetic field on the plasma electron density $n_e/n_0$ at $t=64.7$~ps ($z=19.4$~mm): (a) without the field and (b) with $\Omega=1.1$. The color scale is clipped at $n_e/n_0=6$. The cyan dashed ellipse marks the $2\sigma$ outline of the electron driver (centered at $\zeta\simeq0$, not plotted). Without the field, sheath electrons converge to a sharp on-axis caustic at the back of the first bucket. With the field, the returning flux forms a broad on-axis column farther behind the driver, followed by a second image, consistent with the gyration mechanism. The dashed box outlines the region analyzed in Fig.~\ref{fig:Comparison_w_wo_Bfield}.
}
\end{figure*}

Here we propose another solution to the ``positron problem" in nonlinear plasma wakefields by imposing a strong axial magnetic field in the plasma to increase the phase space for positron acceleration, which is simple and straight-forward. Axial magnetic fields have been studied for beam stabilization~\cite{Su1987,Galyamin2013,Balakirev2001}, as guiding elements for positron beams~\cite{Xu2020}, and for conserving canonical angular momentum (CAM) and suppressing electron injection~\cite{Bulanov2013,Rassou2015,Zhao2019} in laser wakefield accelerators (LWFA).
Wakefields in magnetized electron-positron plasmas have been analyzed in the pulsar context~\cite{Mofiz1989}. Magnetized plasma wakefield acceleration in relativistic astrophysical outflows from AGN (Active Galactic Nuclei) or GRB (Gamma Ray Burst) has been proposed as the mechanism for ultra-high energy cosmic ray production \cite{ChenTajimaTakahashi2002,Chang2009}. In our preliminary two-dimensional study of the concept of magnetized plasma wakefield in the blowout regime for positron acceleration~\cite{ChenLiu2026}, we reported that the magnetic field does regularize plasma wakefields, but without establishing the mechanism, the positioning law, and the witness transport.

We will show that for an electron driver, the CAM physics that suppresses electron trapping in LWFA has an opposite consequence for positrons. The field does not prevent the expelled electrons from returning toward the axis; rather, it re-times their return. Instead of collapsing promptly into a sharp density spike at the back of the first bucket, electrons gyrate out through the bubble periphery and re-converge on the cyclotron timescale as a broad, smooth, finite-radius column. In the co-moving frame, this forms a quasi-achromatic gyro image. The resulting on-axis electron column focuses positrons exactly where the wake accelerates them. Unlike previous remedies, this scheme requires only a conventional electron driver and a uniform plasma, together with the external solenoid. Using quasi-3D and full-3D simulations, we show that this self-imaging structure transports a positron witness bunch with $92\%$ charge capture rate and a $100$--$150$~MeV energy gain over a $6$~cm stage, which corresponds to an acceleration gradient of $1.6$--$2.5$~GeV/m.

We simulate this configuration using the quasi-3D particle-in-cell code Smilei~\cite{Derouillat2018,Zemzemi2020}. An electron driver ($\gamma_d=1000$, Gaussian $1/e$ half-widths $\sigma_z=16\;\mu$m and $\sigma_r=50\;\mu$m, peak density $n_b=2.5\,n_0$) propagates in a uniform plasma ($n_0=10^{16}\,\mathrm{cm^{-3}}$, $\lambda_p=334\;\mu$m, $E_0=m_ec\omega_p/e=9.6$~GV/m) immersed in a uniform axial field $B_z=35.3$~T ($\Omega\equiv\omega_c/\omega_p=1.1$), on a grid of $\Delta r=4\;\mu$m by $\Delta z=1\;\mu$m. The driver is idealized as cold, with zero emittance and energy spread; a realistic driver phase space is left to future work. Results are examined through resolution tests, cross-code comparison with WarpX~\cite{Vay2018,WarpX2607}, and full-3D simulations; numerical details and benchmark tables are given in the companion paper~\cite{Companion}.

It seems natural to wonder whether the same scheme may do even better if  applied to plasma wakefields driven by positron beams. The clock itself has no preferred sign of driver charge; the driver modifies the spacing of the images through the wake potential, and sets the phase they are anchored to and the quality of the band they form. In our exploratory scans, the resulting quality favors the electron driver. A controlled transport comparison of the two is left to future work. From here on, we will concentrate on the electron-beam-driven case only.

\begin{figure*}[t]
\includegraphics[width=\textwidth]{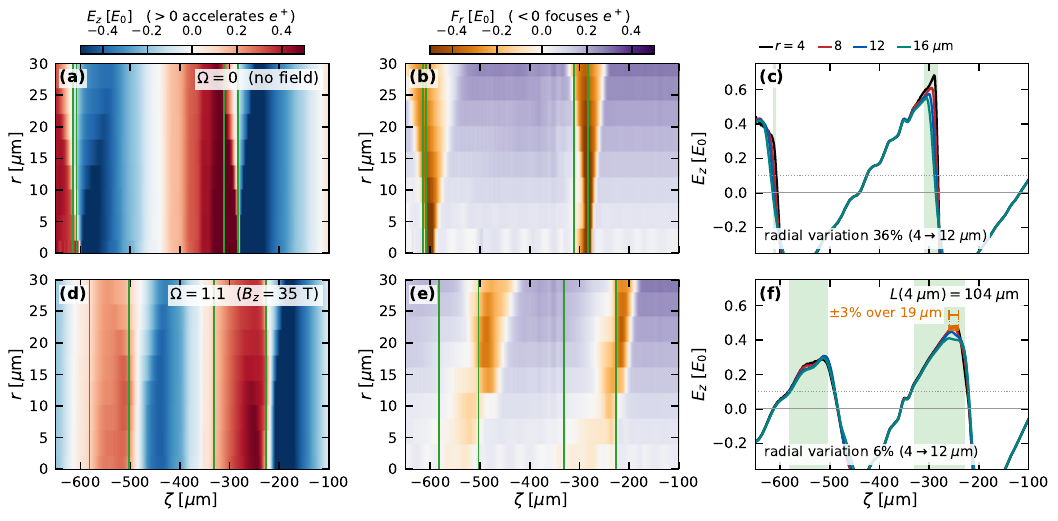}
\caption{\label{fig:Comparison_w_wo_Bfield}
Comparison of the wakefields inside the boxed region of Fig.~\ref{fig:bubble}, smoothed over $11\;\mu$m in $\zeta$. The top row (a--c) shows the unmagnetized case ($\Omega=0$), and the bottom row (d--f) shows the magnetized case ($\Omega=1.1$). The columns display the accelerating field $E_z$, the radial force $F_r$, and $E_z(\zeta)$ sampled at radii $r=4$--$16\;\mu$m. Green lines and shading indicate the usable islands for a $4\;\mu$m acceptance radius. Without the field, the islands are short and the longest exhibits strong radial variation ($36\%$ across $r=4\to12\;\mu$m). At $\Omega=1.1$, the longest island forms a highly uniform band ($6\%$ radial variation) spanning $104\;\mu$m, with a $19\;\mu$m segment flat to within $\pm3\%$ (at $1\;\mu$m radial resolution the magnetized advantage in median usable length is a factor of $2.7$~\cite{Companion}).
}
\end{figure*}

Figure~\ref{fig:bubble} illustrates the impact of the magnetic field. Without it, sheath electrons return to the axis as a sharp caustic (peaking at $42\,n_0$) at the back of the first bucket. Embedded in a magnetic field with $\Omega=1.1$, they gyrate through the bubble periphery and re-converge on the cyclotron timescale to form a broad, smooth column ($5.9\,n_0$).

This structural change creates an ideal environment for positrons. Figure~\ref{fig:Comparison_w_wo_Bfield} compares the ``usable'' accelerating intervals for a positron bunch, defined as regions where the longitudinal field is strong ($E_z > 0.10\,E_0$) and uniform (within $20\%$), while the radial force ($F_r\equiv E_r-B_\theta$) is focusing ($F_r<0$) up to a given acceptance radius $R$. At $R=6\;\mu$m, the magnetized wake provides a continuous usable interval of $94\;\mu$m, more than four times longer than the matched unmagnetized case ($22\;\mu$m). At larger radii the contrast widens: at a $12\;\mu$m acceptance radius the unmagnetized wake retains only $6\;\mu$m of usable length, against $57\;\mu$m with the field. These figures compare the two wakes during band formation; the settled band is narrower, with the full acceptance-radius sweep and resolution tests given in the companion paper~\cite{Companion}.

Two invariants of a magnetized wake govern this re-imaging process. First, the quasi-static invariant sets the longitudinal spacing. For electrons initially at rest ahead of the driver, the wake conserves $\gamma-p_z=1+\psi(\zeta,r)$~\cite{Mora1997}, where $\psi=\phi-a_z$ is the wake pseudo-potential (potentials normalized to $m_ec^2/e$, momenta to $m_ec$, lengths to $c/\omega_p$, and times to $\omega_p^{-1}$). The uniform axial field $B_z$ does not enter $\psi$, leaving this invariant unchanged. Electrons expelled by the driver around a co-moving position $\zeta_c$ into the small-potential periphery ($\psi\to0$) undergo relativistic cyclotron rotation with a period $T=2\pi\gamma/\Omega$. In the co-moving frame, neglecting transverse wake fields and variations in $\psi$, their backward slippage over one gyro-period is
\begin{equation}
\Delta\zeta=(1-\beta_z)\,cT=\frac{2\pi}{\Omega}\,(\gamma-p_z)
=\frac{\lambda_p}{\Omega}\,(1+\psi).
\label{eq:imaging}
\end{equation}
In the periphery, this slippage simplifies to $\lambda_c\equiv\lambda_p/\Omega$, which is independent of the individual electron's energy. This acts as a universal clock in this limit: electrons expelled together arrive back at the same co-moving positions, forming successive gyro images at $\zeta_N=\zeta_c-N\lambda_c$ ($N=1,2,\dots$).

Second, conservation of canonical angular momentum, $\ell=p_\theta r-\Omega r^2/2$, sets the transverse scale. An electron starting at rest at radius $r_0>0$ acquires $|\ell|=\Omega r_0^2/2$, which prevents it from reaching the axis at finite momentum ($r_{\min}\approx\Omega r_0^2/2p_\perp$ for $r_{\min}\ll r_0$). Consequently, returning electrons are individually held off-axis, helping transform what would be a sharp singularity into a finite-radius column. While this CAM-induced off-axis return suppresses electron trapping in LWFA~\cite{Bulanov2013,Rassou2015,Zhao2019}, it provides precisely the extended, smooth negative-charge structure required to focus positrons.

\begin{figure}[htbp]
\includegraphics[width=\columnwidth]{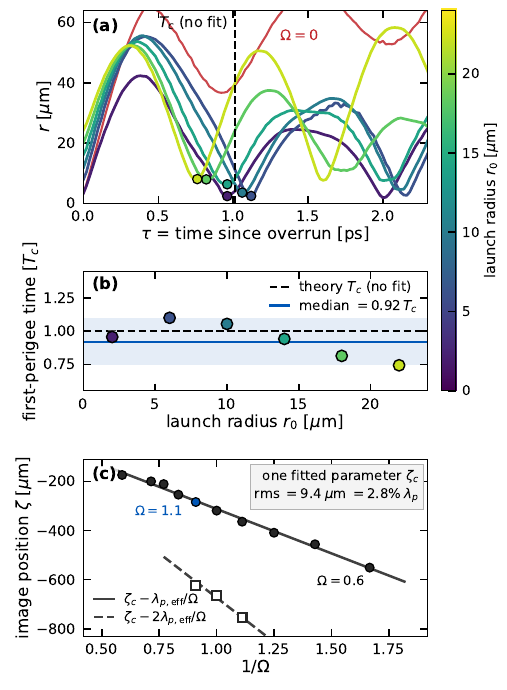}
\caption{\label{fig:gyro_clock}
Tests of the gyro clock mechanism. (a) Median radius of tracer electrons launched from six initial radii (cadence of $0.05\,T_c$) versus time since overrun by the driver. Circles mark the first perigee; the dashed line is $T_c=2\pi/(\Omega\omega_p)$. The red line shows the $\Omega=0$ control median, which has no comparable return minimum over the plotted interval. (b) First-perigee time for different launch radii, showing launch-radius dependence about a grand median of $0.92\,T_c$. (c) The positioning law: on-axis density-peak position versus $1/\Omega$ for the first (filled) and second (open) images. The blue marker is the working point ($\Omega=1.1$). The solid line is Eq.~(\ref{eq:zone}) drawn with $\lambda_{p,\mathrm{eff}}=361\;\mu$m; peaks are selected with reference to this law~\cite{Companion}.}
\end{figure}

Figure~\ref{fig:gyro_clock} supports this physical picture. Tracing the electron orbits reveals a median first perigee at $0.92\,T_c$, with return times varying across launch radii. The finite pseudo-potential experienced while crossing the non-neutral regions makes the imaging quasi-achromatic rather than exact, while a separate positron-driver scan calibrates the effective spatial period at $\lambda_{p,\mathrm{eff}}=361\;\mu$m~\cite{Companion}. With this calibration fixed and only the expulsion phase $\zeta_c$ fitted, the image positions follow
\begin{equation}
\zeta_N=\zeta_c-N\,\lambda_{p,\mathrm{eff}}/\Omega ,\qquad N=1,2,\dots
\label{eq:zone}
\end{equation}
This positioning law describes the selected density-peak locations across a wide scan of magnetic fields [Fig.~\ref{fig:gyro_clock}(c)]. The clock is written by the magnetic field, while the anchor $\zeta_c$ and aberration are written by the driver.
The working point is not finely tuned: the quality interval opens near $\Om\simeq0.8$, with the largest sampled width--field product at $\Om=1.1$. The imaging law tracks $1/\Om$ across the full scan, and the magnetic-field and injection tolerances are quantified in the companion paper~\cite{Companion}.

While Eq.~(\ref{eq:zone}) predicts the initial formation position of the first image (around $\zeta=-280\;\mu$m for $\Omega=1.1$), the usable band is not static. During the driver's early transient evolution (compression and re-expansion), the band migrates forward and eventually locks into a stable co-moving window between $\zeta=-234$ and $-212\;\mu$m after $z\approx44$~mm. To test longitudinal transport, we initialize a low-charge positron witness bunch ($\gamma_w=5000$, $\varepsilon_n=50$~mm\,mrad, $\sigma_r=7\;\mu$m, $\sigma_z=4\;\mu$m, $0.1$~pC) at this empirically settled position ($\zeta_0=-227\;\mu$m) at $t=0$, ensuring the witness is present when the band locks.

\begin{figure}[htbp]
\includegraphics[width=\columnwidth]{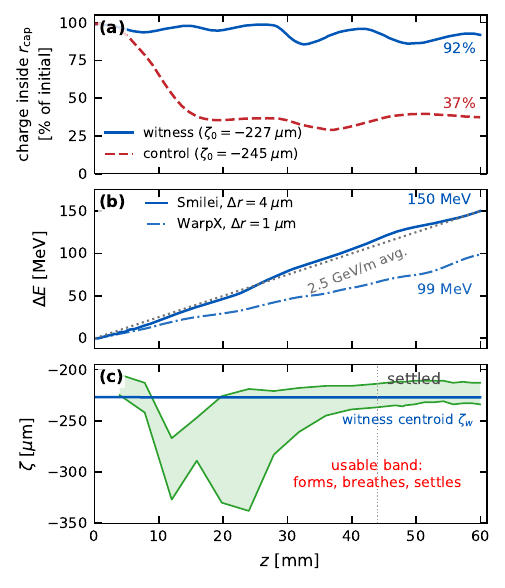}
\caption{\label{fig:witness}
Witness transport over a $200$-ps quasi-3D simulation at $\Omega=1.1$. (a) Captured charge (within $r<21.5\;\mu$m) for a witness bunch injected at the settled position ($\zeta_0=-227\;\mu$m) versus a control bunch injected $18\;\mu$m farther back. (b) Energy gain of the captured ensemble on the baseline grid ($150$~MeV) and the high-resolution WarpX benchmark ($1\;\mu$m radial cells, $99$~MeV). The dotted line indicates a $2.5$~GeV/m average. (c) Evolution of the usable band and the witness centroid $\zeta_w$. The band initially breathes and then settles around the bunch after $z\approx44$~mm.
}
\end{figure}

Figure~\ref{fig:witness} tracks the witness over a $60$-mm quasi-3D simulation. The centroid is nearly kinematically pinned, so that the stable band settles onto the bunch rather than the bunch slipping out. The witness concludes the stage with $92\%$ of its initial charge captured within the diagnostic $r<21.5\;\mu$m aperture, gaining $150$~MeV (an average gradient of $2.5$~GeV/m); the finest WarpX grid gives $99$~MeV ($1.6$~GeV/m), with gain convergence still unresolved~\cite{Companion}. Full-3D simulations of the entire $60$-mm stage corroborate this transport: with the witness placed at the same structure-relative position (in 3D the accelerating band sits ${\sim}25\;\mu$m farther behind the driver), the bunch is captured at $85\%$ and accelerated through the end of the stage. Detailed benchmarks between the quasi-3D and full-3D frameworks are provided in the companion paper~\cite{Companion}.

The witness emittance grows by a factor of $1.7$ and then saturates within the stage~\cite{Companion}, so this configuration serves as a robust single-stage demonstrator.
The structure also survives beam loading. Repeating the simulation with only the witness charge varied, and evaluating all cases at a common $99$~ps ($30$~mm) epoch, we found that the capture rate remains at $92$--$94\%$ and the emittance grows by only $13\%$ relative to the $10$~pC case up to $200$~pC (where the bunch density reaches $40\,n_0$). Beyond this, the bunch overloads the wake, and the energy gain changes sign between the $200$ and $400$~pC runs. Evaluated using the dimensionless luminosity per unit power of Ref.~\cite{Cao2024}, this sampled $200$~pC optimum reaches $\tilde L_P\simeq0.07$. This is nearly six orders of magnitude above the $0.1$~pC probe value and roughly one order of magnitude below the best published positron schemes compared in Ref.~\cite{Cao2024}. The remaining performance gap depends on both the beam parameters and the evolving wake structure. Specifically, the short accelerating band accommodates a short bunch, $k_p\sigma_z=0.075$. This short bunch limits the plasma response during its passage, while the extraction efficiency is of the same order as $k_p\sigma_z$~\cite{Cao2024,Companion}. Furthermore, the output emittance is currently nearly three orders of magnitude above the ${\sim}0.1$~mm\,mrad of collider designs~\cite{Cao2024}, and simply injecting a higher-quality beam does not fully resolve this: on a $1\;\mu$m grid, with the witness injected before the wake has formed, a five-fold reduction of the initial emittance, accompanied by a smaller rms radius, reduces the final emittance by only a factor of $2.2$~\cite{Companion}. Improving these parameters is left to future work. Canonical angular momentum exchange at the solenoid fringes (Busch's theorem~\cite{Busch1926,Reiser2008}) gives a small estimated witness kick, while the driver entrance state can shift the early image~\cite{Companion}.

The clearest path forward is scaling to higher energies. Increasing the driver energy to $\gamma_d=5000$ ($2.6$~GeV) extends the demonstrated transport, with a similar band structure settling at $z\approx117$~mm instead of $44$~mm. A positron witness transported through that longer stage confirms this directly: over $200$~mm it retains a $93\%$ capture rate and gains $419$~MeV, at a gradient averaged over $117$--$200$~mm within $6\%$ of the $\gamma_d=1000$ value over $40$--$60$~mm. It is still gaining at $1.8$~GeV/m over the final $24$~mm, so the settled window extends for at least $83$~mm~\cite{Companion}. The settling length itself grew by a factor of $2.7$ for a fivefold increase in driver energy, somewhat faster than the $\sqrt{\gamma_d}$ of the betatron scale~\cite{Companion}; whether this trend extends to a $10$-GeV-class driver remains to be tested.

All results here scale naturally with density at fixed normalized beam and plasma parameters, maintaining $B\simeq32\,\Omega\,\sqrt{n_0/10^{16}\,\mathrm{cm^{-3}}}$~T. Operating near $10^{16}\,\mathrm{cm^{-3}}$ anchors the requirement at $35$~T, a static field strength already achievable in laboratory magnets combining superconducting coils with resistive background fields~\cite{Hahn2019}. While accelerating positrons in the wake of an electron driver is a known concept~\cite{Lotov2007,Wang2008}, the available accelerating regions are typically short, highly nonuniform, and challenging to use. By employing an axial magnetic field, this highly restricted wake is reshaped into a stable, self-imaging accelerating structure. The field governs the cyclotron clock that refocuses plasma electrons, providing simultaneous focusing and multi-GeV/m acceleration for a positron bunch positioned at the settled phase of the first gyro image.

\section*{Data availability}
The data that support the findings of this work are available from the authors upon reasonable request. On publication the input decks, the analysis scripts and the reduced data behind every figure will be deposited in a public archive with a DOI.

\begin{acknowledgments}
We thank the National Center for High-performance Computing (NCHC), Taiwan, for providing the Taiwania 3 supercomputer and computing resources.
PC appreciates the support by Taiwan's National Science and Technology Council (NSTC) under funding accounts: 115-2112-M-002-014- and 115-2221-E-002-189-MY3.
Work at SLAC is supported by the U.S. Department of Energy under contract DE-AC02-76SF00515.
\end{acknowledgments}

\bibliography{references}

\begin{thebibliography}{40}%
\makeatletter
\providecommand \@ifxundefined [1]{%
 \@ifx{#1\undefined}
}%
\providecommand \@ifnum [1]{%
 \ifnum #1\expandafter \@firstoftwo
 \else \expandafter \@secondoftwo
 \fi
}%
\providecommand \@ifx [1]{%
 \ifx #1\expandafter \@firstoftwo
 \else \expandafter \@secondoftwo
 \fi
}%
\providecommand \natexlab [1]{#1}%
\providecommand \enquote  [1]{``#1''}%
\providecommand \bibnamefont  [1]{#1}%
\providecommand \bibfnamefont [1]{#1}%
\providecommand \citenamefont [1]{#1}%
\providecommand \href@noop [0]{\@secondoftwo}%
\providecommand \href [0]{\begingroup \@sanitize@url \@href}%
\providecommand \@href[1]{\@@startlink{#1}\@@href}%
\providecommand \@@href[1]{\endgroup#1\@@endlink}%
\providecommand \@sanitize@url [0]{\catcode `\\12\catcode `\$12\catcode
  `\&12\catcode `\#12\catcode `\^12\catcode `\_12\catcode `\%12\relax}%
\providecommand \@@startlink[1]{}%
\providecommand \@@endlink[0]{}%
\providecommand \url  [0]{\begingroup\@sanitize@url \@url }%
\providecommand \@url [1]{\endgroup\@href {#1}{\urlprefix }}%
\providecommand \urlprefix  [0]{URL }%
\providecommand \Eprint [0]{\href }%
\providecommand \doibase [0]{https://doi.org/}%
\providecommand \selectlanguage [0]{\@gobble}%
\providecommand \bibinfo  [0]{\@secondoftwo}%
\providecommand \bibfield  [0]{\@secondoftwo}%
\providecommand \translation [1]{[#1]}%
\providecommand \BibitemOpen [0]{}%
\providecommand \bibitemStop [0]{}%
\providecommand \bibitemNoStop [0]{.\EOS\space}%
\providecommand \EOS [0]{\spacefactor3000\relax}%
\providecommand \BibitemShut  [1]{\csname bibitem#1\endcsname}%
\let\auto@bib@innerbib\@empty
\bibitem [{\citenamefont {Tajima}\ and\ \citenamefont
  {Dawson}(1979)}]{Tajima1979}%
  \BibitemOpen
  \bibfield  {author} {\bibinfo {author} {\bibfnamefont {T.}~\bibnamefont
  {Tajima}}\ and\ \bibinfo {author} {\bibfnamefont {J.~M.}\ \bibnamefont
  {Dawson}},\ }\bibfield  {title} {\bibinfo {title} {Laser electron
  accelerator},\ }\href@noop {} {\bibfield  {journal} {\bibinfo  {journal}
  {Phys. Rev. Lett.}\ }\textbf {\bibinfo {volume} {43}},\ \bibinfo {pages}
  {267} (\bibinfo {year} {1979})}\BibitemShut {NoStop}%
\bibitem [{\citenamefont {Chen}\ \emph {et~al.}(1985)\citenamefont {Chen},
  \citenamefont {Dawson}, \citenamefont {Huff},\ and\ \citenamefont
  {Katsouleas}}]{Chen1985}%
  \BibitemOpen
  \bibfield  {author} {\bibinfo {author} {\bibfnamefont {P.}~\bibnamefont
  {Chen}}, \bibinfo {author} {\bibfnamefont {J.~M.}\ \bibnamefont {Dawson}},
  \bibinfo {author} {\bibfnamefont {R.~W.}\ \bibnamefont {Huff}},\ and\
  \bibinfo {author} {\bibfnamefont {T.}~\bibnamefont {Katsouleas}},\ }\bibfield
   {title} {\bibinfo {title} {Acceleration of electrons by the interaction of a
  bunched electron beam with a plasma},\ }\href@noop {} {\bibfield  {journal}
  {\bibinfo  {journal} {Phys. Rev. Lett.}\ }\textbf {\bibinfo {volume} {54}},\
  \bibinfo {pages} {693} (\bibinfo {year} {1985})}\BibitemShut {NoStop}%
\bibitem [{\citenamefont {Rosenzweig}\ \emph {et~al.}(1991)\citenamefont
  {Rosenzweig}, \citenamefont {Breizman}, \citenamefont {Katsouleas},\ and\
  \citenamefont {Su}}]{Rosenzweig1991}%
  \BibitemOpen
  \bibfield  {author} {\bibinfo {author} {\bibfnamefont {J.~B.}\ \bibnamefont
  {Rosenzweig}}, \bibinfo {author} {\bibfnamefont {B.}~\bibnamefont
  {Breizman}}, \bibinfo {author} {\bibfnamefont {T.}~\bibnamefont
  {Katsouleas}},\ and\ \bibinfo {author} {\bibfnamefont {J.~J.}\ \bibnamefont
  {Su}},\ }\bibfield  {title} {\bibinfo {title} {Acceleration and focusing of
  electrons in two-dimensional nonlinear plasma wake fields},\ }\href@noop {}
  {\bibfield  {journal} {\bibinfo  {journal} {Phys. Rev. A}\ }\textbf {\bibinfo
  {volume} {44}},\ \bibinfo {pages} {R6189} (\bibinfo {year}
  {1991})}\BibitemShut {NoStop}%
\bibitem [{\citenamefont {Lu}\ \emph {et~al.}(2006)\citenamefont {Lu},
  \citenamefont {Huang}, \citenamefont {Zhou}, \citenamefont {Mori},\ and\
  \citenamefont {Katsouleas}}]{Lu2006}%
  \BibitemOpen
  \bibfield  {author} {\bibinfo {author} {\bibfnamefont {W.}~\bibnamefont
  {Lu}}, \bibinfo {author} {\bibfnamefont {C.}~\bibnamefont {Huang}}, \bibinfo
  {author} {\bibfnamefont {M.}~\bibnamefont {Zhou}}, \bibinfo {author}
  {\bibfnamefont {W.~B.}\ \bibnamefont {Mori}},\ and\ \bibinfo {author}
  {\bibfnamefont {T.}~\bibnamefont {Katsouleas}},\ }\bibfield  {title}
  {\bibinfo {title} {Nonlinear theory for relativistic plasma wakefields in the
  blowout regime},\ }\href@noop {} {\bibfield  {journal} {\bibinfo  {journal}
  {Phys. Rev. Lett.}\ }\textbf {\bibinfo {volume} {96}},\ \bibinfo {pages}
  {165002} (\bibinfo {year} {2006})}\BibitemShut {NoStop}%
\bibitem [{\citenamefont {Cao}\ \emph {et~al.}(2024)\citenamefont {Cao},
  \citenamefont {Lindstr{\o}m}, \citenamefont {Adli}, \citenamefont {Corde},\
  and\ \citenamefont {Gessner}}]{Cao2024}%
  \BibitemOpen
  \bibfield  {author} {\bibinfo {author} {\bibfnamefont {G.~J.}\ \bibnamefont
  {Cao}}, \bibinfo {author} {\bibfnamefont {C.~A.}\ \bibnamefont
  {Lindstr{\o}m}}, \bibinfo {author} {\bibfnamefont {E.}~\bibnamefont {Adli}},
  \bibinfo {author} {\bibfnamefont {S.}~\bibnamefont {Corde}},\ and\ \bibinfo
  {author} {\bibfnamefont {S.}~\bibnamefont {Gessner}},\ }\bibfield  {title}
  {\bibinfo {title} {Positron acceleration in plasma wakefields},\ }\href@noop
  {} {\bibfield  {journal} {\bibinfo  {journal} {Phys. Rev. Accel. Beams}\
  }\textbf {\bibinfo {volume} {27}},\ \bibinfo {pages} {034801} (\bibinfo
  {year} {2024})}\BibitemShut {NoStop}%
\bibitem [{\citenamefont {Joshi}\ \emph {et~al.}(2025)\citenamefont {Joshi},
  \citenamefont {Mori},\ and\ \citenamefont {Hogan}}]{Joshi2025}%
  \BibitemOpen
  \bibfield  {author} {\bibinfo {author} {\bibfnamefont {C.}~\bibnamefont
  {Joshi}}, \bibinfo {author} {\bibfnamefont {W.~B.}\ \bibnamefont {Mori}},\
  and\ \bibinfo {author} {\bibfnamefont {M.~J.}\ \bibnamefont {Hogan}},\
  }\bibfield  {title} {\bibinfo {title} {The positron arm of a plasma-based
  linear collider},\ }\href@noop {} {\bibfield  {journal} {\bibinfo  {journal}
  {Nat. Phys.}\ }\textbf {\bibinfo {volume} {21}},\ \bibinfo {pages} {885}
  (\bibinfo {year} {2025})}\BibitemShut {NoStop}%
\bibitem [{\citenamefont {Chen}\ and\ \citenamefont {Liu}(2026)}]{ChenLiu2026}%
  \BibitemOpen
  \bibfield  {author} {\bibinfo {author} {\bibfnamefont {P.}~\bibnamefont
  {Chen}}\ and\ \bibinfo {author} {\bibfnamefont {Y.-K.}\ \bibnamefont {Liu}},\
  }\bibfield  {title} {\bibinfo {title} {Plasma wakefield: from accelerators to
  black holes},\ }\href {https://doi.org/10.1007/s41614-026-00217-x} {\bibfield
   {journal} {\bibinfo  {journal} {Reviews of Modern Plasma Physics}\ }\textbf
  {\bibinfo {volume} {10}},\ \bibinfo {pages} {7} (\bibinfo {year}
  {2026})}\BibitemShut {NoStop}%
\bibitem [{\citenamefont {Gessner}\ \emph {et~al.}(2016)\citenamefont {Gessner}
  \emph {et~al.}}]{Gessner2016}%
  \BibitemOpen
  \bibfield  {author} {\bibinfo {author} {\bibfnamefont {S.}~\bibnamefont
  {Gessner}} \emph {et~al.},\ }\bibfield  {title} {\bibinfo {title}
  {Demonstration of a positron beam-driven hollow channel plasma wakefield
  accelerator},\ }\href@noop {} {\bibfield  {journal} {\bibinfo  {journal}
  {Nat. Commun.}\ }\textbf {\bibinfo {volume} {7}},\ \bibinfo {pages} {11785}
  (\bibinfo {year} {2016})}\BibitemShut {NoStop}%
\bibitem [{\citenamefont {Lindstr{\o}m}\ \emph {et~al.}(2018)\citenamefont
  {Lindstr{\o}m} \emph {et~al.}}]{Lindstrom2018}%
  \BibitemOpen
  \bibfield  {author} {\bibinfo {author} {\bibfnamefont {C.~A.}\ \bibnamefont
  {Lindstr{\o}m}} \emph {et~al.},\ }\bibfield  {title} {\bibinfo {title}
  {Measurement of transverse wakefields induced by a misaligned positron bunch
  in a hollow channel plasma accelerator},\ }\href
  {https://doi.org/10.1103/PhysRevLett.120.124802} {\bibfield  {journal}
  {\bibinfo  {journal} {Phys. Rev. Lett.}\ }\textbf {\bibinfo {volume} {120}},\
  \bibinfo {pages} {124802} (\bibinfo {year} {2018})}\BibitemShut {NoStop}%
\bibitem [{\citenamefont {Silva}\ \emph {et~al.}(2021)\citenamefont {Silva}
  \emph {et~al.}}]{Silva2021}%
  \BibitemOpen
  \bibfield  {author} {\bibinfo {author} {\bibfnamefont {T.}~\bibnamefont
  {Silva}} \emph {et~al.},\ }\bibfield  {title} {\bibinfo {title} {Stable
  positron acceleration in thin, warm, hollow plasma channels},\ }\href@noop {}
  {\bibfield  {journal} {\bibinfo  {journal} {Phys. Rev. Lett.}\ }\textbf
  {\bibinfo {volume} {127}},\ \bibinfo {pages} {104801} (\bibinfo {year}
  {2021})}\BibitemShut {NoStop}%
\bibitem [{\citenamefont {Diederichs}\ \emph {et~al.}(2019)\citenamefont
  {Diederichs}, \citenamefont {Mehrling}, \citenamefont {Benedetti},
  \citenamefont {Schroeder}, \citenamefont {Knetsch}, \citenamefont {Esarey},\
  and\ \citenamefont {Osterhoff}}]{Diederichs2019}%
  \BibitemOpen
  \bibfield  {author} {\bibinfo {author} {\bibfnamefont {S.}~\bibnamefont
  {Diederichs}}, \bibinfo {author} {\bibfnamefont {T.~J.}\ \bibnamefont
  {Mehrling}}, \bibinfo {author} {\bibfnamefont {C.}~\bibnamefont {Benedetti}},
  \bibinfo {author} {\bibfnamefont {C.~B.}\ \bibnamefont {Schroeder}}, \bibinfo
  {author} {\bibfnamefont {A.}~\bibnamefont {Knetsch}}, \bibinfo {author}
  {\bibfnamefont {E.}~\bibnamefont {Esarey}},\ and\ \bibinfo {author}
  {\bibfnamefont {J.}~\bibnamefont {Osterhoff}},\ }\bibfield  {title} {\bibinfo
  {title} {Positron transport and acceleration in beam-driven plasma wakefield
  accelerators using plasma columns},\ }\href@noop {} {\bibfield  {journal}
  {\bibinfo  {journal} {Phys. Rev. Accel. Beams}\ }\textbf {\bibinfo {volume}
  {22}},\ \bibinfo {pages} {081301} (\bibinfo {year} {2019})}\BibitemShut
  {NoStop}%
\bibitem [{\citenamefont {Diederichs}\ \emph {et~al.}(2020)\citenamefont
  {Diederichs}, \citenamefont {Benedetti}, \citenamefont {Esarey},
  \citenamefont {Osterhoff},\ and\ \citenamefont {Schroeder}}]{Diederichs2020}%
  \BibitemOpen
  \bibfield  {author} {\bibinfo {author} {\bibfnamefont {S.}~\bibnamefont
  {Diederichs}}, \bibinfo {author} {\bibfnamefont {C.}~\bibnamefont
  {Benedetti}}, \bibinfo {author} {\bibfnamefont {E.}~\bibnamefont {Esarey}},
  \bibinfo {author} {\bibfnamefont {J.}~\bibnamefont {Osterhoff}},\ and\
  \bibinfo {author} {\bibfnamefont {C.~B.}\ \bibnamefont {Schroeder}},\
  }\bibfield  {title} {\bibinfo {title} {High-quality positron acceleration in
  beam-driven plasma accelerators},\ }\href
  {https://doi.org/10.1103/PhysRevAccelBeams.23.121301} {\bibfield  {journal}
  {\bibinfo  {journal} {Phys. Rev. Accel. Beams}\ }\textbf {\bibinfo {volume}
  {23}},\ \bibinfo {pages} {121301} (\bibinfo {year} {2020})}\BibitemShut
  {NoStop}%
\bibitem [{\citenamefont {Diederichs}\ \emph {et~al.}(2022)\citenamefont
  {Diederichs}, \citenamefont {Benedetti}, \citenamefont {Th{\'e}venet},
  \citenamefont {Esarey}, \citenamefont {Osterhoff},\ and\ \citenamefont
  {Schroeder}}]{Diederichs2022}%
  \BibitemOpen
  \bibfield  {author} {\bibinfo {author} {\bibfnamefont {S.}~\bibnamefont
  {Diederichs}}, \bibinfo {author} {\bibfnamefont {C.}~\bibnamefont
  {Benedetti}}, \bibinfo {author} {\bibfnamefont {M.}~\bibnamefont
  {Th{\'e}venet}}, \bibinfo {author} {\bibfnamefont {E.}~\bibnamefont
  {Esarey}}, \bibinfo {author} {\bibfnamefont {J.}~\bibnamefont {Osterhoff}},\
  and\ \bibinfo {author} {\bibfnamefont {C.~B.}\ \bibnamefont {Schroeder}},\
  }\bibfield  {title} {\bibinfo {title} {Self-stabilizing positron acceleration
  in a plasma column},\ }\href@noop {} {\bibfield  {journal} {\bibinfo
  {journal} {Phys. Rev. Accel. Beams}\ }\textbf {\bibinfo {volume} {25}},\
  \bibinfo {pages} {091304} (\bibinfo {year} {2022})}\BibitemShut {NoStop}%
\bibitem [{\citenamefont {Wang}\ \emph {et~al.}(2021)\citenamefont {Wang},
  \citenamefont {Khudik},\ and\ \citenamefont {Shvets}}]{Wang2021}%
  \BibitemOpen
  \bibfield  {author} {\bibinfo {author} {\bibfnamefont {T.}~\bibnamefont
  {Wang}}, \bibinfo {author} {\bibfnamefont {V.}~\bibnamefont {Khudik}},\ and\
  \bibinfo {author} {\bibfnamefont {G.}~\bibnamefont {Shvets}},\ }\href@noop {}
  {\bibinfo {title} {Positron acceleration in an elongated bubble regime}}
  (\bibinfo {year} {2021}),\ \Eprint {https://arxiv.org/abs/2110.10290}
  {arXiv:2110.10290 [physics.acc-ph]} \BibitemShut {NoStop}%
\bibitem [{\citenamefont {Jain}\ \emph {et~al.}(2015)\citenamefont {Jain},
  \citenamefont {Antonsen},\ and\ \citenamefont {Palastro}}]{Jain2015}%
  \BibitemOpen
  \bibfield  {author} {\bibinfo {author} {\bibfnamefont {N.}~\bibnamefont
  {Jain}}, \bibinfo {author} {\bibfnamefont {T.~M.}\ \bibnamefont {Antonsen},
  \bibfnamefont {Jr.}},\ and\ \bibinfo {author} {\bibfnamefont {J.~P.}\
  \bibnamefont {Palastro}},\ }\bibfield  {title} {\bibinfo {title} {Positron
  acceleration by plasma wakefields driven by a hollow electron beam},\
  }\href@noop {} {\bibfield  {journal} {\bibinfo  {journal} {Phys. Rev. Lett.}\
  }\textbf {\bibinfo {volume} {115}},\ \bibinfo {pages} {195001} (\bibinfo
  {year} {2015})}\BibitemShut {NoStop}%
\bibitem [{\citenamefont {Vieira}\ and\ \citenamefont {Mendon{\c
  c}a}(2014)}]{Vieira2014}%
  \BibitemOpen
  \bibfield  {author} {\bibinfo {author} {\bibfnamefont {J.}~\bibnamefont
  {Vieira}}\ and\ \bibinfo {author} {\bibfnamefont {J.~T.}\ \bibnamefont
  {Mendon{\c c}a}},\ }\bibfield  {title} {\bibinfo {title} {Nonlinear laser
  driven donut wakefields for positron and electron acceleration},\ }\href@noop
  {} {\bibfield  {journal} {\bibinfo  {journal} {Phys. Rev. Lett.}\ }\textbf
  {\bibinfo {volume} {112}},\ \bibinfo {pages} {215001} (\bibinfo {year}
  {2014})}\BibitemShut {NoStop}%
\bibitem [{\citenamefont {Corde}\ \emph {et~al.}(2015)\citenamefont {Corde}
  \emph {et~al.}}]{Corde2015}%
  \BibitemOpen
  \bibfield  {author} {\bibinfo {author} {\bibfnamefont {S.}~\bibnamefont
  {Corde}} \emph {et~al.},\ }\bibfield  {title} {\bibinfo {title}
  {Multi-gigaelectronvolt acceleration of positrons in a self-loaded plasma
  wakefield},\ }\href@noop {} {\bibfield  {journal} {\bibinfo  {journal}
  {Nature}\ }\textbf {\bibinfo {volume} {524}},\ \bibinfo {pages} {442}
  (\bibinfo {year} {2015})}\BibitemShut {NoStop}%
\bibitem [{\citenamefont {Doche}\ \emph {et~al.}(2017)\citenamefont {Doche}
  \emph {et~al.}}]{Doche2017}%
  \BibitemOpen
  \bibfield  {author} {\bibinfo {author} {\bibfnamefont {A.}~\bibnamefont
  {Doche}} \emph {et~al.},\ }\bibfield  {title} {\bibinfo {title} {Acceleration
  of a trailing positron bunch in a plasma wakefield accelerator},\ }\href@noop
  {} {\bibfield  {journal} {\bibinfo  {journal} {Sci. Rep.}\ }\textbf {\bibinfo
  {volume} {7}},\ \bibinfo {pages} {14180} (\bibinfo {year}
  {2017})}\BibitemShut {NoStop}%
\bibitem [{\citenamefont {Zhou}\ \emph {et~al.}(2025)\citenamefont {Zhou},
  \citenamefont {Ding}, \citenamefont {An}, \citenamefont {Su}, \citenamefont
  {Hua}, \citenamefont {Li}, \citenamefont {Mori}, \citenamefont {Joshi},\ and\
  \citenamefont {Lu}}]{Zhou2025}%
  \BibitemOpen
  \bibfield  {author} {\bibinfo {author} {\bibfnamefont {S.}~\bibnamefont
  {Zhou}}, \bibinfo {author} {\bibfnamefont {S.}~\bibnamefont {Ding}}, \bibinfo
  {author} {\bibfnamefont {W.}~\bibnamefont {An}}, \bibinfo {author}
  {\bibfnamefont {Q.}~\bibnamefont {Su}}, \bibinfo {author} {\bibfnamefont
  {J.}~\bibnamefont {Hua}}, \bibinfo {author} {\bibfnamefont {F.}~\bibnamefont
  {Li}}, \bibinfo {author} {\bibfnamefont {W.~B.}\ \bibnamefont {Mori}},
  \bibinfo {author} {\bibfnamefont {C.}~\bibnamefont {Joshi}},\ and\ \bibinfo
  {author} {\bibfnamefont {W.}~\bibnamefont {Lu}},\ }\bibfield  {title}
  {\bibinfo {title} {Positron beam loading and acceleration in the blowout
  regime of a plasma wakefield accelerator},\ }\href@noop {} {\bibfield
  {journal} {\bibinfo  {journal} {Research}\ }\textbf {\bibinfo {volume} {8}},\
  \bibinfo {pages} {0878} (\bibinfo {year} {2025})}\BibitemShut {NoStop}%
\bibitem [{\citenamefont {Su}\ \emph {et~al.}(1987)\citenamefont {Su},
  \citenamefont {Katsouleas}, \citenamefont {Dawson}, \citenamefont {Chen},
  \citenamefont {Jones},\ and\ \citenamefont {Keinigs}}]{Su1987}%
  \BibitemOpen
  \bibfield  {author} {\bibinfo {author} {\bibfnamefont {J.~J.}\ \bibnamefont
  {Su}}, \bibinfo {author} {\bibfnamefont {T.}~\bibnamefont {Katsouleas}},
  \bibinfo {author} {\bibfnamefont {J.~M.}\ \bibnamefont {Dawson}}, \bibinfo
  {author} {\bibfnamefont {P.}~\bibnamefont {Chen}}, \bibinfo {author}
  {\bibfnamefont {M.}~\bibnamefont {Jones}},\ and\ \bibinfo {author}
  {\bibfnamefont {R.}~\bibnamefont {Keinigs}},\ }\bibfield  {title} {\bibinfo
  {title} {Stability of the driving bunch in the plasma wakefield
  accelerator},\ }\href {https://doi.org/10.1109/TPS.1987.4316684} {\bibfield
  {journal} {\bibinfo  {journal} {IEEE Trans. Plasma Sci.}\ }\textbf {\bibinfo
  {volume} {15}},\ \bibinfo {pages} {192} (\bibinfo {year} {1987})}\BibitemShut
  {NoStop}%
\bibitem [{\citenamefont {Galyamin}\ \emph {et~al.}(2013)\citenamefont
  {Galyamin}, \citenamefont {Kapshtan},\ and\ \citenamefont
  {Tyukhtin}}]{Galyamin2013}%
  \BibitemOpen
  \bibfield  {author} {\bibinfo {author} {\bibfnamefont {S.~N.}\ \bibnamefont
  {Galyamin}}, \bibinfo {author} {\bibfnamefont {D.~Y.}\ \bibnamefont
  {Kapshtan}},\ and\ \bibinfo {author} {\bibfnamefont {A.~V.}\ \bibnamefont
  {Tyukhtin}},\ }\bibfield  {title} {\bibinfo {title} {Electromagnetic field of
  a charge moving in a cold magnetized plasma},\ }\href
  {https://doi.org/10.1103/PhysRevE.87.013109} {\bibfield  {journal} {\bibinfo
  {journal} {Phys. Rev. E}\ }\textbf {\bibinfo {volume} {87}},\ \bibinfo
  {pages} {013109} (\bibinfo {year} {2013})}\BibitemShut {NoStop}%
\bibitem [{\citenamefont {Balakirev}\ \emph {et~al.}(2001)\citenamefont
  {Balakirev}, \citenamefont {Karas'}, \citenamefont {Karas'},\ and\
  \citenamefont {Levchenko}}]{Balakirev2001}%
  \BibitemOpen
  \bibfield  {author} {\bibinfo {author} {\bibfnamefont {V.~A.}\ \bibnamefont
  {Balakirev}}, \bibinfo {author} {\bibfnamefont {V.~I.}\ \bibnamefont
  {Karas'}}, \bibinfo {author} {\bibfnamefont {I.~V.}\ \bibnamefont {Karas'}},\
  and\ \bibinfo {author} {\bibfnamefont {V.~D.}\ \bibnamefont {Levchenko}},\
  }\bibfield  {title} {\bibinfo {title} {Plasma wake-field excitation by
  relativistic electron bunches and charged particle acceleration in the
  presence of external magnetic field},\ }\href
  {https://doi.org/10.1017/S0263034601194061} {\bibfield  {journal} {\bibinfo
  {journal} {Laser Part. Beams}\ }\textbf {\bibinfo {volume} {19}},\ \bibinfo
  {pages} {597} (\bibinfo {year} {2001})}\BibitemShut {NoStop}%
\bibitem [{\citenamefont {Xu}\ \emph {et~al.}(2020)\citenamefont {Xu},
  \citenamefont {Yi}, \citenamefont {Shen}, \citenamefont {Xu}, \citenamefont
  {Ji}, \citenamefont {Xu}, \citenamefont {Zhang}, \citenamefont {Li},\ and\
  \citenamefont {Xu}}]{Xu2020}%
  \BibitemOpen
  \bibfield  {author} {\bibinfo {author} {\bibfnamefont {Z.}~\bibnamefont
  {Xu}}, \bibinfo {author} {\bibfnamefont {L.}~\bibnamefont {Yi}}, \bibinfo
  {author} {\bibfnamefont {B.}~\bibnamefont {Shen}}, \bibinfo {author}
  {\bibfnamefont {J.}~\bibnamefont {Xu}}, \bibinfo {author} {\bibfnamefont
  {L.}~\bibnamefont {Ji}}, \bibinfo {author} {\bibfnamefont {T.}~\bibnamefont
  {Xu}}, \bibinfo {author} {\bibfnamefont {L.}~\bibnamefont {Zhang}}, \bibinfo
  {author} {\bibfnamefont {S.}~\bibnamefont {Li}},\ and\ \bibinfo {author}
  {\bibfnamefont {Z.}~\bibnamefont {Xu}},\ }\bibfield  {title} {\bibinfo
  {title} {Driving positron beam acceleration with coherent transition
  radiation},\ }\href@noop {} {\bibfield  {journal} {\bibinfo  {journal}
  {Commun. Phys.}\ }\textbf {\bibinfo {volume} {3}},\ \bibinfo {pages} {191}
  (\bibinfo {year} {2020})}\BibitemShut {NoStop}%
\bibitem [{\citenamefont {Bulanov}\ \emph {et~al.}(2013)\citenamefont
  {Bulanov}, \citenamefont {Esirkepov}, \citenamefont {Kando}, \citenamefont
  {Koga}, \citenamefont {Hosokai}, \citenamefont {Zhidkov},\ and\ \citenamefont
  {Kodama}}]{Bulanov2013}%
  \BibitemOpen
  \bibfield  {author} {\bibinfo {author} {\bibfnamefont {S.~V.}\ \bibnamefont
  {Bulanov}}, \bibinfo {author} {\bibfnamefont {T.~Z.}\ \bibnamefont
  {Esirkepov}}, \bibinfo {author} {\bibfnamefont {M.}~\bibnamefont {Kando}},
  \bibinfo {author} {\bibfnamefont {J.~K.}\ \bibnamefont {Koga}}, \bibinfo
  {author} {\bibfnamefont {T.}~\bibnamefont {Hosokai}}, \bibinfo {author}
  {\bibfnamefont {A.~G.}\ \bibnamefont {Zhidkov}},\ and\ \bibinfo {author}
  {\bibfnamefont {R.}~\bibnamefont {Kodama}},\ }\bibfield  {title} {\bibinfo
  {title} {Nonlinear plasma wave in magnetized plasmas},\ }\href@noop {}
  {\bibfield  {journal} {\bibinfo  {journal} {Phys. Plasmas}\ }\textbf
  {\bibinfo {volume} {20}},\ \bibinfo {pages} {083113} (\bibinfo {year}
  {2013})}\BibitemShut {NoStop}%
\bibitem [{\citenamefont {Rassou}\ \emph {et~al.}(2015)\citenamefont {Rassou},
  \citenamefont {Bourdier},\ and\ \citenamefont {Drouin}}]{Rassou2015}%
  \BibitemOpen
  \bibfield  {author} {\bibinfo {author} {\bibfnamefont {S.}~\bibnamefont
  {Rassou}}, \bibinfo {author} {\bibfnamefont {A.}~\bibnamefont {Bourdier}},\
  and\ \bibinfo {author} {\bibfnamefont {M.}~\bibnamefont {Drouin}},\
  }\bibfield  {title} {\bibinfo {title} {Influence of a strong longitudinal
  magnetic field on laser wakefield acceleration},\ }\href@noop {} {\bibfield
  {journal} {\bibinfo  {journal} {Phys. Plasmas}\ }\textbf {\bibinfo {volume}
  {22}},\ \bibinfo {pages} {073104} (\bibinfo {year} {2015})}\BibitemShut
  {NoStop}%
\bibitem [{\citenamefont {Zhao}\ \emph {et~al.}(2019)\citenamefont {Zhao},
  \citenamefont {Weng}, \citenamefont {Chen}, \citenamefont {Zeng},
  \citenamefont {Hidding}, \citenamefont {Jaroszynski}, \citenamefont
  {Assmann},\ and\ \citenamefont {Sheng}}]{Zhao2019}%
  \BibitemOpen
  \bibfield  {author} {\bibinfo {author} {\bibfnamefont {Q.}~\bibnamefont
  {Zhao}}, \bibinfo {author} {\bibfnamefont {S.~M.}\ \bibnamefont {Weng}},
  \bibinfo {author} {\bibfnamefont {M.}~\bibnamefont {Chen}}, \bibinfo {author}
  {\bibfnamefont {M.}~\bibnamefont {Zeng}}, \bibinfo {author} {\bibfnamefont
  {B.}~\bibnamefont {Hidding}}, \bibinfo {author} {\bibfnamefont {D.~A.}\
  \bibnamefont {Jaroszynski}}, \bibinfo {author} {\bibfnamefont
  {R.}~\bibnamefont {Assmann}},\ and\ \bibinfo {author} {\bibfnamefont {Z.~M.}\
  \bibnamefont {Sheng}},\ }\bibfield  {title} {\bibinfo {title}
  {Sub-femtosecond electron bunches in laser wakefield acceleration via
  injection suppression with a magnetic field},\ }\href@noop {} {\bibfield
  {journal} {\bibinfo  {journal} {Plasma Phys. Control. Fusion}\ }\textbf
  {\bibinfo {volume} {61}},\ \bibinfo {pages} {085015} (\bibinfo {year}
  {2019})}\BibitemShut {NoStop}%
\bibitem [{\citenamefont {Mofiz}(1989)}]{Mofiz1989}%
  \BibitemOpen
  \bibfield  {author} {\bibinfo {author} {\bibfnamefont {U.~A.}\ \bibnamefont
  {Mofiz}},\ }\bibfield  {title} {\bibinfo {title} {Wake-field accelerator in a
  magnetized electron-positron plasma},\ }\href
  {https://doi.org/10.1103/PhysRevA.40.6752} {\bibfield  {journal} {\bibinfo
  {journal} {Phys. Rev. A}\ }\textbf {\bibinfo {volume} {40}},\ \bibinfo
  {pages} {6752} (\bibinfo {year} {1989})}\BibitemShut {NoStop}%
\bibitem [{\citenamefont {Chen}\ \emph {et~al.}(2002)\citenamefont {Chen},
  \citenamefont {Tajima},\ and\ \citenamefont
  {Takahashi}}]{ChenTajimaTakahashi2002}%
  \BibitemOpen
  \bibfield  {author} {\bibinfo {author} {\bibfnamefont {P.}~\bibnamefont
  {Chen}}, \bibinfo {author} {\bibfnamefont {T.}~\bibnamefont {Tajima}},\ and\
  \bibinfo {author} {\bibfnamefont {Y.}~\bibnamefont {Takahashi}},\ }\bibfield
  {title} {\bibinfo {title} {Plasma wakefield acceleration for ultrahigh-energy
  cosmic rays},\ }\href {https://doi.org/10.1103/PhysRevLett.89.161101}
  {\bibfield  {journal} {\bibinfo  {journal} {Phys. Rev. Lett.}\ }\textbf
  {\bibinfo {volume} {89}},\ \bibinfo {pages} {161101} (\bibinfo {year}
  {2002})}\BibitemShut {NoStop}%
\bibitem [{\citenamefont {Chang}\ \emph {et~al.}(2009)\citenamefont {Chang},
  \citenamefont {Chen}, \citenamefont {Lin}, \citenamefont {Noble},\ and\
  \citenamefont {Sydora}}]{Chang2009}%
  \BibitemOpen
  \bibfield  {author} {\bibinfo {author} {\bibfnamefont {F.-Y.}\ \bibnamefont
  {Chang}}, \bibinfo {author} {\bibfnamefont {P.}~\bibnamefont {Chen}},
  \bibinfo {author} {\bibfnamefont {G.-L.}\ \bibnamefont {Lin}}, \bibinfo
  {author} {\bibfnamefont {R.}~\bibnamefont {Noble}},\ and\ \bibinfo {author}
  {\bibfnamefont {R.}~\bibnamefont {Sydora}},\ }\bibfield  {title} {\bibinfo
  {title} {Magnetowave induced plasma wakefield acceleration for ultrahigh
  energy cosmic rays},\ }\href {https://doi.org/10.1103/PhysRevLett.102.111101}
  {\bibfield  {journal} {\bibinfo  {journal} {Phys. Rev. Lett.}\ }\textbf
  {\bibinfo {volume} {102}},\ \bibinfo {pages} {111101} (\bibinfo {year}
  {2009})}\BibitemShut {NoStop}%
\bibitem [{\citenamefont {Derouillat}\ \emph {et~al.}(2018)\citenamefont
  {Derouillat}, \citenamefont {Beck}, \citenamefont {P{\'e}rez}, \citenamefont
  {Vinci}, \citenamefont {Chiaramello}, \citenamefont {Grassi}, \citenamefont
  {Fl{\'e}}, \citenamefont {Bouchard}, \citenamefont {Plotnikov}, \citenamefont
  {Aunai}, \citenamefont {Dargent}, \citenamefont {Riconda},\ and\
  \citenamefont {Grech}}]{Derouillat2018}%
  \BibitemOpen
  \bibfield  {author} {\bibinfo {author} {\bibfnamefont {J.}~\bibnamefont
  {Derouillat}}, \bibinfo {author} {\bibfnamefont {A.}~\bibnamefont {Beck}},
  \bibinfo {author} {\bibfnamefont {F.}~\bibnamefont {P{\'e}rez}}, \bibinfo
  {author} {\bibfnamefont {T.}~\bibnamefont {Vinci}}, \bibinfo {author}
  {\bibfnamefont {M.}~\bibnamefont {Chiaramello}}, \bibinfo {author}
  {\bibfnamefont {A.}~\bibnamefont {Grassi}}, \bibinfo {author} {\bibfnamefont
  {M.}~\bibnamefont {Fl{\'e}}}, \bibinfo {author} {\bibfnamefont
  {G.}~\bibnamefont {Bouchard}}, \bibinfo {author} {\bibfnamefont
  {I.}~\bibnamefont {Plotnikov}}, \bibinfo {author} {\bibfnamefont
  {N.}~\bibnamefont {Aunai}}, \bibinfo {author} {\bibfnamefont
  {J.}~\bibnamefont {Dargent}}, \bibinfo {author} {\bibfnamefont
  {C.}~\bibnamefont {Riconda}},\ and\ \bibinfo {author} {\bibfnamefont
  {M.}~\bibnamefont {Grech}},\ }\bibfield  {title} {\bibinfo {title} {Smilei: A
  collaborative, open-source, multi-purpose particle-in-cell code for plasma
  simulation},\ }\href@noop {} {\bibfield  {journal} {\bibinfo  {journal}
  {Comput. Phys. Commun.}\ }\textbf {\bibinfo {volume} {222}},\ \bibinfo
  {pages} {351} (\bibinfo {year} {2018})}\BibitemShut {NoStop}%
\bibitem [{\citenamefont {Zemzemi}\ \emph {et~al.}(2020)\citenamefont
  {Zemzemi}, \citenamefont {Massimo},\ and\ \citenamefont
  {Beck}}]{Zemzemi2020}%
  \BibitemOpen
  \bibfield  {author} {\bibinfo {author} {\bibfnamefont {I.}~\bibnamefont
  {Zemzemi}}, \bibinfo {author} {\bibfnamefont {F.}~\bibnamefont {Massimo}},\
  and\ \bibinfo {author} {\bibfnamefont {A.}~\bibnamefont {Beck}},\ }\bibfield
  {title} {\bibinfo {title} {Azimuthal decomposition study of a realistic laser
  profile for efficient modeling of laser wakefield acceleration},\ }\href
  {https://doi.org/10.1088/1742-6596/1596/1/012054} {\bibfield  {journal}
  {\bibinfo  {journal} {J. Phys.: Conf. Ser.}\ }\textbf {\bibinfo {volume}
  {1596}},\ \bibinfo {pages} {012054} (\bibinfo {year} {2020})}\BibitemShut
  {NoStop}%
\bibitem [{\citenamefont {Vay}\ \emph {et~al.}(2018)\citenamefont {Vay},
  \citenamefont {Almgren}, \citenamefont {Bell}, \citenamefont {Ge},
  \citenamefont {Grote}, \citenamefont {Hogan}, \citenamefont {Kononenko},
  \citenamefont {Lehe}, \citenamefont {Myers}, \citenamefont {Ng},
  \citenamefont {Park}, \citenamefont {Ryne}, \citenamefont {Shapoval},
  \citenamefont {Th{\'e}venet},\ and\ \citenamefont {Zhang}}]{Vay2018}%
  \BibitemOpen
  \bibfield  {author} {\bibinfo {author} {\bibfnamefont {J.-L.}\ \bibnamefont
  {Vay}}, \bibinfo {author} {\bibfnamefont {A.}~\bibnamefont {Almgren}},
  \bibinfo {author} {\bibfnamefont {J.}~\bibnamefont {Bell}}, \bibinfo {author}
  {\bibfnamefont {L.}~\bibnamefont {Ge}}, \bibinfo {author} {\bibfnamefont
  {D.~P.}\ \bibnamefont {Grote}}, \bibinfo {author} {\bibfnamefont
  {M.}~\bibnamefont {Hogan}}, \bibinfo {author} {\bibfnamefont
  {O.}~\bibnamefont {Kononenko}}, \bibinfo {author} {\bibfnamefont
  {R.}~\bibnamefont {Lehe}}, \bibinfo {author} {\bibfnamefont {A.}~\bibnamefont
  {Myers}}, \bibinfo {author} {\bibfnamefont {C.}~\bibnamefont {Ng}}, \bibinfo
  {author} {\bibfnamefont {J.}~\bibnamefont {Park}}, \bibinfo {author}
  {\bibfnamefont {R.}~\bibnamefont {Ryne}}, \bibinfo {author} {\bibfnamefont
  {O.}~\bibnamefont {Shapoval}}, \bibinfo {author} {\bibfnamefont
  {M.}~\bibnamefont {Th{\'e}venet}},\ and\ \bibinfo {author} {\bibfnamefont
  {W.}~\bibnamefont {Zhang}},\ }\bibfield  {title} {\bibinfo {title} {Warp-{X}:
  A new exascale computing platform for beam-plasma simulations},\ }\href
  {https://doi.org/10.1016/j.nima.2018.01.035} {\bibfield  {journal} {\bibinfo
  {journal} {Nucl. Instrum. Methods Phys. Res. A}\ }\textbf {\bibinfo {volume}
  {909}},\ \bibinfo {pages} {476} (\bibinfo {year} {2018})}\BibitemShut
  {NoStop}%
\bibitem [{\citenamefont {{The WarpX Development Team}}(2026)}]{WarpX2607}%
  \BibitemOpen
  \bibfield  {author} {\bibinfo {author} {\bibnamefont {{The WarpX Development
  Team}}},\ }\href {https://doi.org/10.5281/zenodo.21268496} {\bibinfo {title}
  {{WarpX}, version 26.07}} (\bibinfo {year} {2026}),\ \bibinfo {note} {rZ
  geometry, FDTD solver}\BibitemShut {NoStop}%
\bibitem [{\citenamefont {Liu}\ \emph {et~al.}(2026)\citenamefont {Liu},
  \citenamefont {Chen}, \citenamefont {Lin}, \citenamefont {Gessner},\ and\
  \citenamefont {Hidding}}]{Companion}%
  \BibitemOpen
  \bibfield  {author} {\bibinfo {author} {\bibfnamefont {Y.-K.}\ \bibnamefont
  {Liu}}, \bibinfo {author} {\bibfnamefont {P.}~\bibnamefont {Chen}}, \bibinfo
  {author} {\bibfnamefont {C.-E.}\ \bibnamefont {Lin}}, \bibinfo {author}
  {\bibfnamefont {S.}~\bibnamefont {Gessner}},\ and\ \bibinfo {author}
  {\bibfnamefont {B.}~\bibnamefont {Hidding}},\ }\href@noop {} {\bibinfo
  {title} {\cptitlerevision{Magnetizing nonlinear plasma wakefields for
  positron acceleration: mechanism and operating limits}}} (\bibinfo {year}
  {2026}),\ \bibinfo {note} {companion paper},\ \Eprint
  {https://arxiv.org/abs/2609.33175} {arXiv:2609.33175 [physics.acc-ph]}
  \BibitemShut {NoStop}%
\bibitem [{\citenamefont {Mora}\ and\ \citenamefont
  {Antonsen}(1997)}]{Mora1997}%
  \BibitemOpen
  \bibfield  {author} {\bibinfo {author} {\bibfnamefont {P.}~\bibnamefont
  {Mora}}\ and\ \bibinfo {author} {\bibfnamefont {T.~M.}\ \bibnamefont
  {Antonsen}, \bibfnamefont {Jr.}},\ }\bibfield  {title} {\bibinfo {title}
  {Kinetic modeling of intense, short laser pulses propagating in tenuous
  plasmas},\ }\href@noop {} {\bibfield  {journal} {\bibinfo  {journal} {Phys.
  Plasmas}\ }\textbf {\bibinfo {volume} {4}},\ \bibinfo {pages} {217} (\bibinfo
  {year} {1997})}\BibitemShut {NoStop}%
\bibitem [{\citenamefont {Busch}(1926)}]{Busch1926}%
  \BibitemOpen
  \bibfield  {author} {\bibinfo {author} {\bibfnamefont {H.}~\bibnamefont
  {Busch}},\ }\bibfield  {title} {\bibinfo {title} {Berechnung der {B}ahn von
  {K}athodenstrahlen im axialsymmetrischen elektromagnetischen {F}elde},\
  }\href@noop {} {\bibfield  {journal} {\bibinfo  {journal} {Ann. Phys.
  (Leipzig)}\ }\textbf {\bibinfo {volume} {386}},\ \bibinfo {pages} {974}
  (\bibinfo {year} {1926})}\BibitemShut {NoStop}%
\bibitem [{\citenamefont {Reiser}(2008)}]{Reiser2008}%
  \BibitemOpen
  \bibfield  {author} {\bibinfo {author} {\bibfnamefont {M.}~\bibnamefont
  {Reiser}},\ }\href@noop {} {\emph {\bibinfo {title} {Theory and Design of
  Charged Particle Beams}}},\ \bibinfo {edition} {2nd}\ ed.\ (\bibinfo
  {publisher} {Wiley-VCH},\ \bibinfo {address} {Weinheim},\ \bibinfo {year}
  {2008})\BibitemShut {NoStop}%
\bibitem [{\citenamefont {Hahn}\ \emph {et~al.}(2019)\citenamefont {Hahn},
  \citenamefont {Kim}, \citenamefont {Kim}, \citenamefont {Hu}, \citenamefont
  {Painter}, \citenamefont {Dixon}, \citenamefont {Kim}, \citenamefont
  {Bhattarai}, \citenamefont {Noguchi}, \citenamefont {Jaroszynski},\ and\
  \citenamefont {Larbalestier}}]{Hahn2019}%
  \BibitemOpen
  \bibfield  {author} {\bibinfo {author} {\bibfnamefont {S.}~\bibnamefont
  {Hahn}}, \bibinfo {author} {\bibfnamefont {K.}~\bibnamefont {Kim}}, \bibinfo
  {author} {\bibfnamefont {K.}~\bibnamefont {Kim}}, \bibinfo {author}
  {\bibfnamefont {X.}~\bibnamefont {Hu}}, \bibinfo {author} {\bibfnamefont
  {T.}~\bibnamefont {Painter}}, \bibinfo {author} {\bibfnamefont
  {I.}~\bibnamefont {Dixon}}, \bibinfo {author} {\bibfnamefont
  {S.}~\bibnamefont {Kim}}, \bibinfo {author} {\bibfnamefont {K.~R.}\
  \bibnamefont {Bhattarai}}, \bibinfo {author} {\bibfnamefont {S.}~\bibnamefont
  {Noguchi}}, \bibinfo {author} {\bibfnamefont {J.}~\bibnamefont
  {Jaroszynski}},\ and\ \bibinfo {author} {\bibfnamefont {D.~C.}\ \bibnamefont
  {Larbalestier}},\ }\bibfield  {title} {\bibinfo {title} {{45.5-tesla}
  direct-current magnetic field generated with a high-temperature
  superconducting magnet},\ }\href@noop {} {\bibfield  {journal} {\bibinfo
  {journal} {Nature}\ }\textbf {\bibinfo {volume} {570}},\ \bibinfo {pages}
  {496} (\bibinfo {year} {2019})}\BibitemShut {NoStop}%
\bibitem [{\citenamefont {Lotov}(2007)}]{Lotov2007}%
  \BibitemOpen
  \bibfield  {author} {\bibinfo {author} {\bibfnamefont {K.~V.}\ \bibnamefont
  {Lotov}},\ }\bibfield  {title} {\bibinfo {title} {Acceleration of positrons
  by electron beam-driven wakefields in a plasma},\ }\href@noop {} {\bibfield
  {journal} {\bibinfo  {journal} {Phys. Plasmas}\ }\textbf {\bibinfo {volume}
  {14}},\ \bibinfo {pages} {023101} (\bibinfo {year} {2007})}\BibitemShut
  {NoStop}%
\bibitem [{\citenamefont {Wang}\ \emph {et~al.}(2008)\citenamefont {Wang},
  \citenamefont {Ischebeck}, \citenamefont {Muggli}, \citenamefont
  {Katsouleas}, \citenamefont {Joshi}, \citenamefont {Mori},\ and\
  \citenamefont {Hogan}}]{Wang2008}%
  \BibitemOpen
  \bibfield  {author} {\bibinfo {author} {\bibfnamefont {X.}~\bibnamefont
  {Wang}}, \bibinfo {author} {\bibfnamefont {R.}~\bibnamefont {Ischebeck}},
  \bibinfo {author} {\bibfnamefont {P.}~\bibnamefont {Muggli}}, \bibinfo
  {author} {\bibfnamefont {T.}~\bibnamefont {Katsouleas}}, \bibinfo {author}
  {\bibfnamefont {C.}~\bibnamefont {Joshi}}, \bibinfo {author} {\bibfnamefont
  {W.~B.}\ \bibnamefont {Mori}},\ and\ \bibinfo {author} {\bibfnamefont
  {M.~J.}\ \bibnamefont {Hogan}},\ }\bibfield  {title} {\bibinfo {title}
  {Positron injection and acceleration on the wake driven by an electron beam
  in a foil-and-gas plasma},\ }\href@noop {} {\bibfield  {journal} {\bibinfo
  {journal} {Phys. Rev. Lett.}\ }\textbf {\bibinfo {volume} {101}},\ \bibinfo
  {pages} {124801} (\bibinfo {year} {2008})}\BibitemShut {NoStop}%
\end{thebibliography}%

\end{document}